# Security Aware Mobile Web Service Provisioning

**Satish Srirama*, Matthias Jarke[1,2], Wolfgang Prinz[1,2] and Kiran Pendyala**
[1]RWTH Aachen, Informatik V
Ahornstr.55, 52056 Aachen, Germany
[2]Fraunhofer FIT
Schloss Birlinghoven, 53754 Sankt Augustin, Germany
{srirama, jarke, pendyala}@i5.informatik.rwth-aachen.de
wolfgang.prinz@fit.fraunhofer.de
*Corresponding author

**Abstract:** Mobile data services in combination with profluent web services are seemingly the path breaking domain in current information research. Effectively, these mobile web services will pave the way for exciting performance and security challenges, the core need-to-be-addressed issues. On security front, though a lot of standardized security specifications and implementations exist for web services in the wired networks, not much has been analysed and standardized in the wireless environments. This paper addresses some of the critical challenges in providing security to the mobile web service domain. We first explore mobile web services and their key security issues, with special focus on provisioning based on a mobile web service provider realized by us. Later we discuss state-of-the-art security awareness in the wired and wireless web services, and finally address the realization of security for the mobile web service provisioning with performance analysis results.

**Biographical notes:** Satish Srirama is a research assistant at Informatik V, RWTH Aachen University and holder of scholarship from GraduationKollegs funded by German Research Foundation (DFG). He received his Masters in Software Systems Engineering from RWTH Aachen University and Bachelors degree in Computer Science and Systems Engineering from Andhra University. His current research focuses on mobile web services, service oriented architectures and mobile community support.

Matthias Jarke is Professor of Information Systems at RWTH Aachen University and Director of the FIT, Fraunhofer Institute for Applied Information Technology in Sankt Augustin, Germany. His research interest is information systems support for cooperative applications in engineering, business, and culture. Jarke served as Chief Editor of Information Systems from 1993-2004 and is currently president of GI, the German Informatics society.

Wolfgang Prinz is Professor for Cooperation Systems at RWTH Aachen and Director and Head of CSCW Research Department at FIT, Fraunhofer Institute for Applied Information Technology in Sankt Augustin, Germany. His research interests are Cooperative Software Systems, Enterprise Content Management, CSCW, Groupware, CommunityWare, Social Computing, Cooperative Knowledge Management, and Pervasive Games.

Kiran Pendyala is a Masters student in Media Informatics from RWTH Aachen University. He is currently doing his master thesis with title "security aspects analysis in mobile web service provisioning". He received his Bachelors degree in Computer Sciences from Bangalore University. His research interests are Ubiquitous Computing, Post Desktop User Interfaces and mobile web services.

## 1 INTRODUCTION

As mobile and wireless applications are growing ubiquitous in conjunction with the fast developing web services, the ability to provide secure and reliable communication in the vulnerable and volatile mobile ad-hoc topologies is vastly becoming necessary. Unfortunately these pervasive networks, where a central administration is not always possible; the security implementation is hard and challenging. Moreover with the advent of easily readable web services, mobile web services, the complexity to realize security increased further. Secure provisioning of mobile web services needs proper identification mechanism, access control, data integrity and confidentiality. It also requires policies and trust relations to be established between users as well as between users and service providers.

Even though a lot of standardized security specifications, protocols and implementations like WS-Security [1], SAML [2] etc., exist for web services in traditional wired networks, not much has been explored and standardized in wireless environments, with feasibility, till date. Some of the reasons for this poor state might be the lack of active commercial data applications due to the limited resources of the mobile terminals like memory, processing capability, and the low transmission rates of the wireless mediums. Our study contributes to this work and tries to bridge this gap, with main focus at realizing some of the existing security standards in the mobile web services domain.

The paper discusses our project "Mobile Web Service Provisioning", with its security issues, challenges and difficulties in inducing the current existing security standards. Breaches specific to web services such as sniffing, tampering, snooping etc. are considered along with general security breaches like man-in-the-middle attacks, denial-of-service attacks, and intrusion. The paper is organized as follows:

In Section 2, we discuss the concept and applications of mobile web service provisioning, whereas section 3 addresses the security challenges in mobile web services. Section 4 describes existing and emerging standards in mobile and web service domains. Section 5 discusses some of the security realization details and their analysis on our Mobile Host. Section 6 concludes the paper with possible future research directions.

## 2 MOBILE WEB SERVICE PROVISIONING

As high end mobile phones and all-ip broadband based mobile networks are fast creeping into the current market, the increase in usage of mobile data services by a roaming user is quite evident. To that effect, lots of organizations have been established, up and running, to accomplish these necessities of this large user base in mobile data services domain [3]. Although today's bulk of the end users still relay on mobiles only for voice calls, the previous argument will augment the usage of other services especially data services very soon. The statistic analysis provided by Idealliance [4] further enhances the point that very near in future; best part of the requirement will end up at a device called high-end mobile phone/handset.

Web services, on the other hand, are an intriguing prospect in the field of current information systems engineering. Web services are a distributed component model, based on Service Oriented Architecture (SOA) [5], with primary motive of achieving loose coupling and cross-platform interoperability across applications. Its big advantage lies in its simplicity in expression, communication and servicing. The core components of web services architecture are Provider, Publisher and Consumer. The Provider develops a service and registers it with the Publisher. The public interfaces of the service are described using WSDL [6]. A prospective Consumer searches for the service at Publisher, gets the WSDL, interprets it and invokes the service from Provider. To mention, quite a few security mechanisms exist for standalone web services.

As the mobile data services and web services are promising enough, the thought of mobile web services is very much on the money within the context. To support the mobile web services, there exists many organisations such as OMA [7], LA [8] on the specification front; some practical data service applications such as OTA (over-the-air provisioning), application handover etc. on the commercial front; and SUN, IBM toolkits [9-10] on the development front. Thus, though this is early stages, we can safely assume that mobile web services are the road ahead.

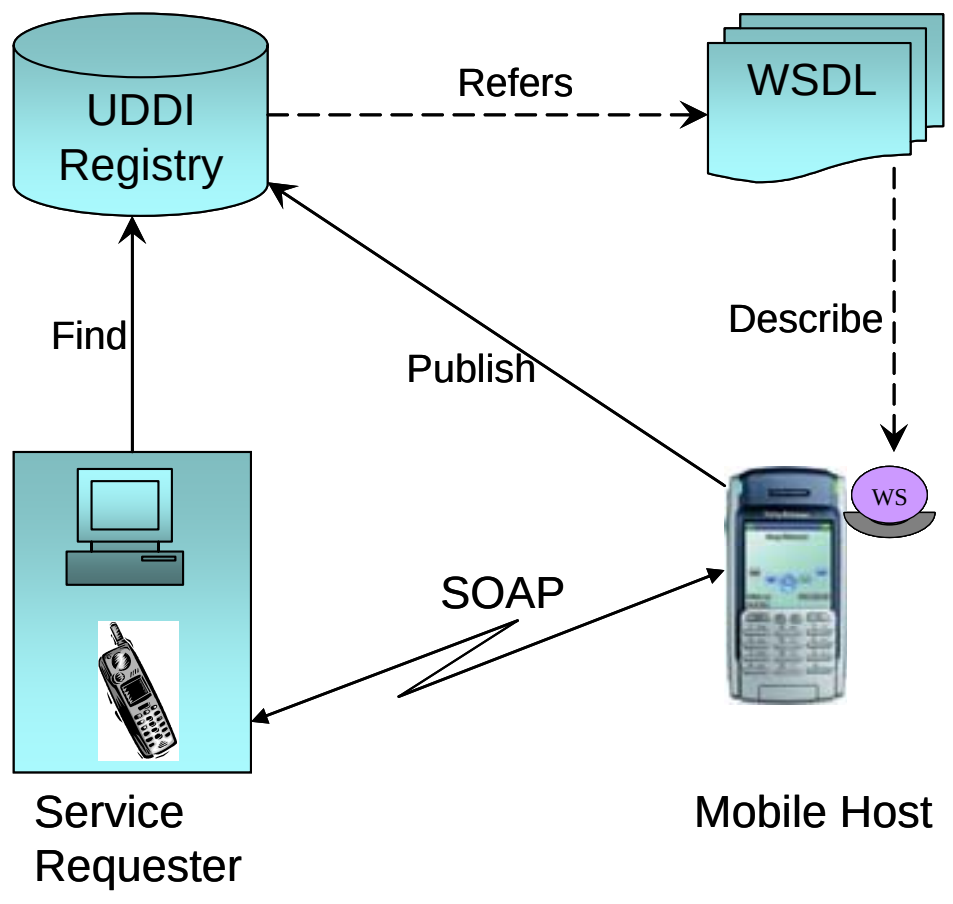

**Figure 1** *Basic mobile web services framework with Mobile Host*

In support of the above argument, a small mobile web service provider ("Mobile Host") [11] has been developed for resource constrained devices. Figure 1 shows the basic mobile web services framework with web services being provided from the Mobile Host. The striking point is that the service provider was possible on one of the earliest high end handset configuration (Sony Ericsson P800). The memory imprint was around ~130KB. The detailed performance analysis conducted with the Mobile Host showed that the processing capability, time frames are very much within the acceptance levels.

In a commercial picture, mobile hosting of web services can exhibit its potential in various ways. Primarily, the cellular phone can act as a multi-user device without additional manual effort on part of the mobile carrier. The approach finds many of its applications in domains like collaborative learning, social systems, mobile community support etc. One can also foresee a strong drive towards P2P architectures in mobile world. Mobile Host thus creates a new trend and lead to manifold opportunities to mobile operators, wireless equipment vendors, third-party application developers, and end users [12].

So the Mobile Host on a latest available configuration, coupled with impressing mobile web services and high speed mobile networks, emphasizes high reality and holds a very exciting prospect. This very fundamental factor is influencing us to further explore this domain. Our current research focuses at QoS aspects of the Mobile Host, with main concern at security implications of this approach.

## 3 SECURITY CHALLENGES FOR MOBILE WEB SERVICES

Once the web service is deployed on the Mobile Host, the service is prone to different types of security breaches like denial-of-service attacks, man-in-the-middle attacks, intrusion, spoofing etc. As web services use message-based technologies for complex transactions across multiple domains, traditional point-to-point security paradigms fall short. Potentially, a web-service message traverses through several intermediaries before it reaches its final destination. Therefore, the need for sophisticated end-to-end message-level security becomes a high priority and is not addressed by existing security technologies.

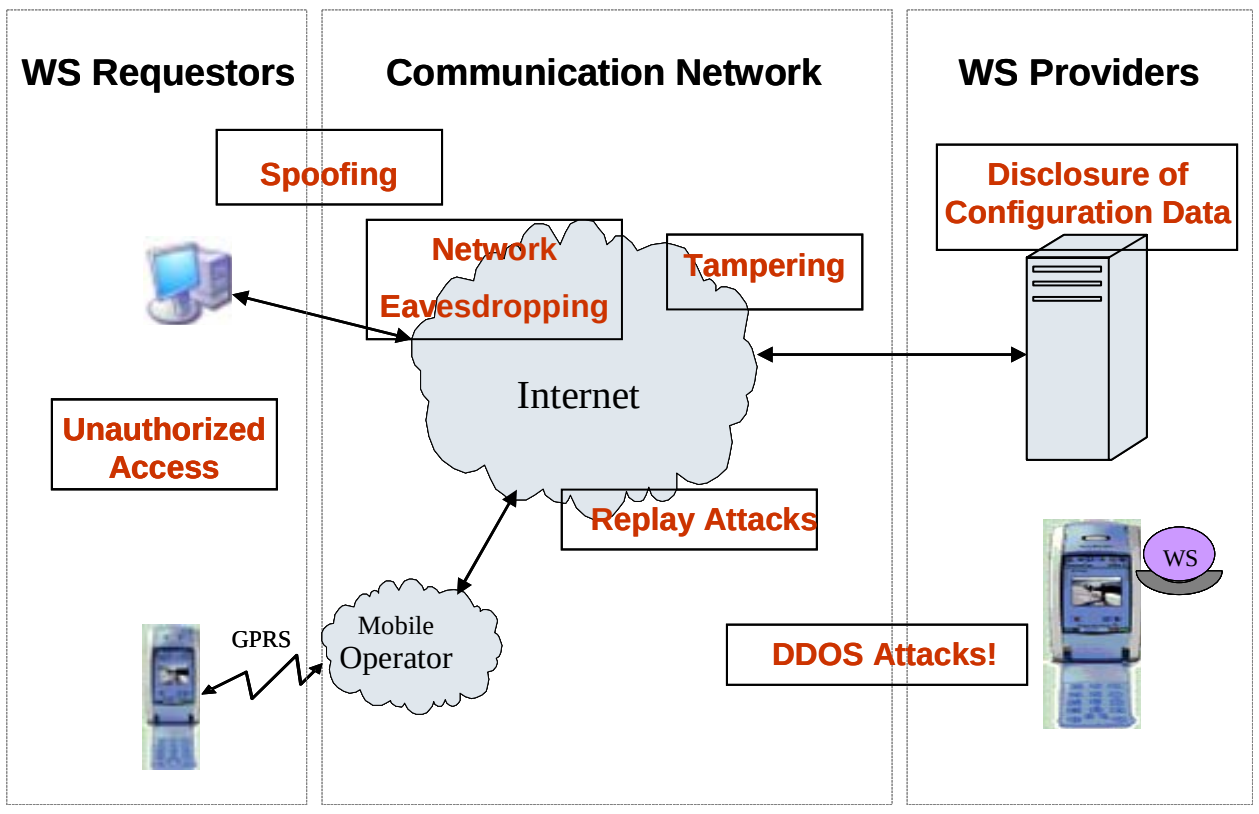

**Figure 2** *Typical security breaches in mobile web services*

Figure 2 depicts some of the typical security breaches in web service and wireless environments. Spoofing is a means of accessing a system with false identity. Proper authentication and authorization principles are to be used to cover spoofing and unauthorized access. Tampering is an act of unauthorized modification of a web service message in network. Network eavesdropping is to monitor traffic for sensitive data such as plaintext passwords by placing sniffers in the middle of the network. Proper encryption and digital signatures can help in avoiding tampering and network eavesdropping attacks.

Generally WSDL files reveal lot of information about a web service and other sensitive information like configuration data of servers. Use authorized access for WSDL downloads to avoid this disclosure. Replaying a valid changed or unchanged message to a web service by impersonating the client is referred as replay attack. The unchanged message replay attack, basic replay attack, can be avoided by using nonce, a cryptographically unique value, with the web service message. Denial-of-service is a process of making a system, server or application unavailable. For each individual service, maintaining and understanding the collection of data can help in protecting it from denial-of-service attacks. But having such a scenario implemented on the resource constrained mobile phones could be impractical. Security policies and high-level access control should help to a certain extent in this regard.

At a minimum, the web service communication should support the basic security requirements as emphasized in figure 3. Secure message transmission is achieved by ensuring confidentiality and data integrity, while authentication and authorization will ensure that the service is accessed only by the trusted service requestors.

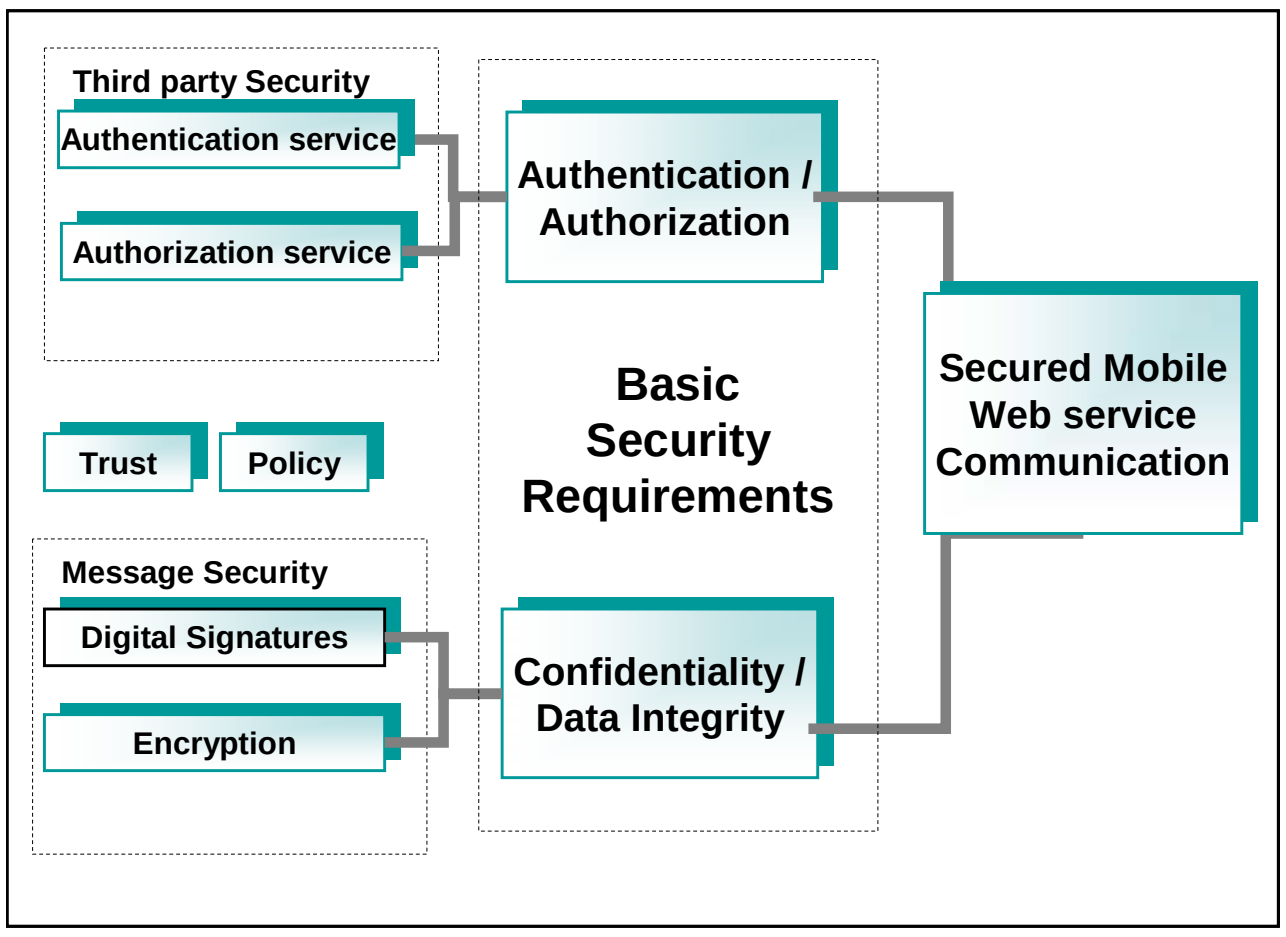

**Figure 3** *Basic security requirements for mobile web services*

## 4 EXISTING/EMERGING STANDARDS

Before considering the realization of security challenges for the mobile web services domain, this section discusses briefly the existing security standards, specifications and some relevant notable projects in web services and wireless domains. Listed below are some of the standard committees and organizations working around web services, wireless domain and their security:

- W3C [13] is primarily responsible for SOAP [14], XML Encryption [15], XML Signature [16] and WSDL standards.
- OASIS [17] is an organization which has larger interest in web service specific standards and it owns primary areas of our interest such as WS-Security and SAML standards.
- Liberty Alliance (LA) group was aimed at providing a framework for interoperable federated identity.
- Open Mobile Alliance (OMA) was formed to develop and promote interoperability for mobile data services.

The web service communication is now-a-days mostly based on SOAP protocol, which in turn exchanges information in XML format. Traditional SOAP protocol from W3C does not exactly provide ways for secure communication. Mainly OASIS standards with the help of XML standards like XML Signature and XML Encryption provides cryptographic protection, and thus help in securing a web service message. The WS-Security specification from OASIS is the core element in web service security realm. It provides ways to add security headers to SOAP envelopes, attach security tokens and credentials to a message, insert a timestamp, sign the messages, and encrypt the message.

Security Assertion Markup Language (SAML) is an extension of WS-Security and specifies the language to exchange identity, attribute and authorization information between various parties involved in web service communication in an interoperable way. This SAML in combination with LA specification framework could help in achieving Single-sign-on. The basic components of LA framework are Principal, Service provider and Identity provider. LA framework's primary provisions are Federation, Single-sign-on and Circle-of-Trust. Federation establishes relationship between any two of the basic LA components. Single-sign-on is a mechanism where the authentication, provided to Principal by the Identity Provider, can be maintained to multiple Service Providers. Circle-of-Trust establishes trust between Service Providers and Identity Providers with agreements upon which Principals can make transactions and exchange information in a seamless and secure way.

In addition, OMA group is concentrating to have a unique specification/framework for mobile data services to achieve interoperability. OMA was formed in June 2002 by nearly 200 companies including the world's leading mobile operators, device and network suppliers, information technology companies and content and service providers. Mobility and roaming are the obvious key characteristics which are hindrances to mobile web service interactions. The current possible mobile web service applications have a number of drawbacks as following. First, the applications should be created through tightly-coupled, costly and close alliances between value-added service providers. Second, they have to be created based on a mixture of mostly propriety models and disparate standards such as WAP, Location, Presence, Identity etc. Furthermore, most of the standards to develop these applications have been devised specifically for the mobile environment from the ground up. All these drawbacks will draw high complexity to deploy, integrate and use these applications and services.

The OMA Web Services Enabler specification [18] is destined to cover all the drawbacks mentioned above and envisioned to support the following mobile web service interactions:

- Server-to-server
- Server-to-mobile terminal
- Mobile terminal-to-server
- Mobile terminal-to-mobile terminal (peer-to-peer)

## 5 SECURITY REALIZATION AND ANALYSIS

As discussed earlier, secure provisioning of mobile web services needs proper message-level security consisting data integrity, confidentiality and end-point access security that constitutes authentication, access control. Since, there exists no approved specific mobile web service standards and lot of propriety interfaces are involved, the security was analyzed on a case-by-case scenario.

### 5.1 Design Model

Figure 4 depicts our architecture to realize the basic security principles for the Mobile Host.

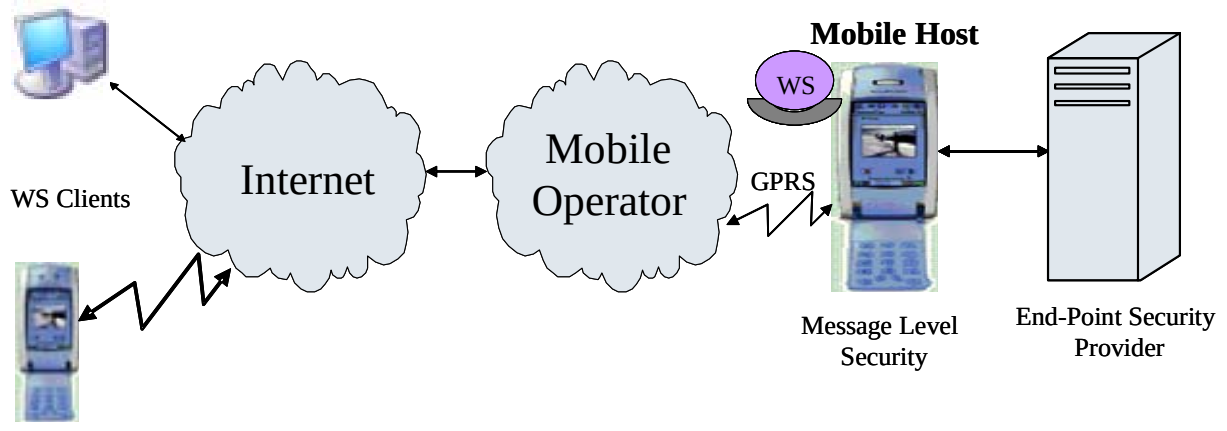

**Figure 4** *Proposed security realization scenario of Mobile Host*

Once a web service is deployed on Mobile Host; any WS client can request for the service. The SOAP message along with the WS-Security information is routed across the internet to the Mobile Host. The message-level security information is extracted and addressed at the Mobile Host while the end-point access security is handled by a third party on behalf of the Mobile Host. Then the corresponding service details are extracted and the service is invoked. The SOAP response is sent back to the client across the same route.

### 5.2 Implementation Model

As far as the realization of the WS-Security for Mobile Host is concerned, we have used J2ME MIDP2.0 [19] for implementation on a Sony Ericsson P910i. The device supports MIDP2.0 with CLDC1.0 [20] configuration. For cryptographic algorithms and digital signers, java based light weight cryptographic API from Bouncy Castle crypto package [21] is used. KSoap2 [22], the java API based on KXml2 [23], is adapted by us according to WS-Security standard and utilized to create the request/response web service messages.

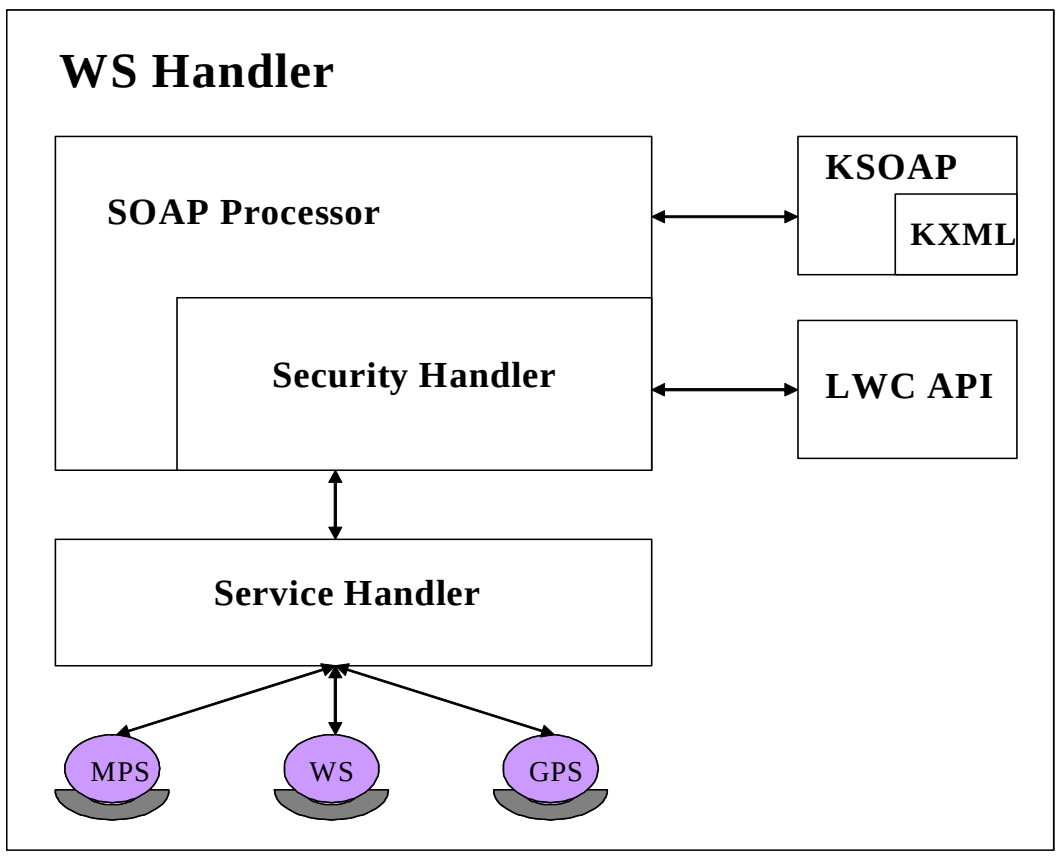

**Figure 5** *Web Service handler of the Mobile Host*

The web service security enabled WS Handler of the Mobile Host is shown in figure 5. The SOAP Processor extracts the SOAP messages from web service requests. The Security Handler does the respective security tasks/checks over the message and transfers decrypted message to the service handler, which extracts the service details and invokes the respective service. Effectively, the handler manages the full message-level security and assists in end-point security.

To realize confidentiality, the message was ciphered with symmetric encryption algorithm and the generated symmetric key is exchanged by means of asymmetric encryption method. The message was tested against various symmetric encryption algorithms including the WS-Security mandatory algorithms, namely, TRIPLEDES [24], AES-128, AES-192 and AES-256 [25]. The PKI algorithm used for key exchange was RSA-V1.5 [26] with 1024 and 2048 bit keys. Upon successful deployment and testing of confidentiality, we considered data integrity on top of confidentiality. The messages were digitally signed and tested against two signer algorithms, namely, DSAwithSHA1 (DSS) [27] and RSAwithSHA1 signature algorithms. Note that, as said earlier, all the algorithms mentioned above have been implemented using java based light weight bouncy castle cryptographic API.

As far as end-point security scenario is concerned, the basic service-level authentication and user-intervened authorization were realized. In the service-level authentication, an authentication service is provided at the mobile web service provider which accepts a username and password and validates the client. Authentication can be password based, Public Key Infrastructure (PKI) based or certificate based. We have considered password based and PKI based authentication due to platform restrictions. The Mobile Host stored the authentication details at the smart phone itself. This posed further problems with the resource constraints of the smart phone, as the authentication information needed extra resources. An alternative for this scenario is provided, where the Mobile Host generates an authentication request to a standalone web service on behalf of the client. The client can then access any service provided by the Mobile Host. Both the authorization-service request and the service request must be generated in a single session. An alternative for the authentication would be the Single-sign-on addressed by SAML and LA.

In the user-intervened authorization, each of the services provided at the Mobile Host can be configured to obtain the providers (person using the Mobile Host) acceptance before providing the respective service to the WS requestor. Critical issues like disapprovals, user being busy and timeouts were also considered. The process was also automated by using an access control mechanism, based on the authentication details. Realization of Single-sign-on, where identity and credentials can be maintained for multiple sessions and parties, is currently under study and our future publications will address this issue.

### 5.3 Performance Model

To analyze the performance of the Mobile Host with the security load, the durations of different activities during the web service invocation cycle are observed. The client initiates the call for the web service and the Mobile Host processes the request, populates the response, and sends response back to the client. The total time taken for this mobile web service invocation ($T_{mwsp}$) constitutes, the time taken by client for constructing valid SOAP message ($T_{cc}$), the time taken to encrypt the message with security information according to WS-Security standard ($T_{reqec}$), the time taken to serialize the encrypted message ($T_{reqs}$), the time taken to transmit the SOAP request to Mobile Host ($T_{reqt}$), the time taken for de-serializing the XML based SOAP request message ($T_{reqd}$), the time taken to decrypt the request message ($T_{reqdc}$), the time taken by the Mobile Host to execute the respective business logic and to populate the response ($T_{process}$), the time taken to encrypt the response message with security information ($T_{resec}$), the time taken for serializing the encrypted response message back to XML data streams ($T_{ress}$), the time taken to transmit the SOAP response back to the client ($T_{rest}$), the time taken to de-serialize the response at the client ($T_{resd}$), the time taken by the client to decrypt the response message ($T_{resdc}$), and lastly the time taken by the client to process the response ($T_{cp}$). The invocation process is shown in Figure 6 and the total time taken for the mobile web service invocation is given in the following equation 1.

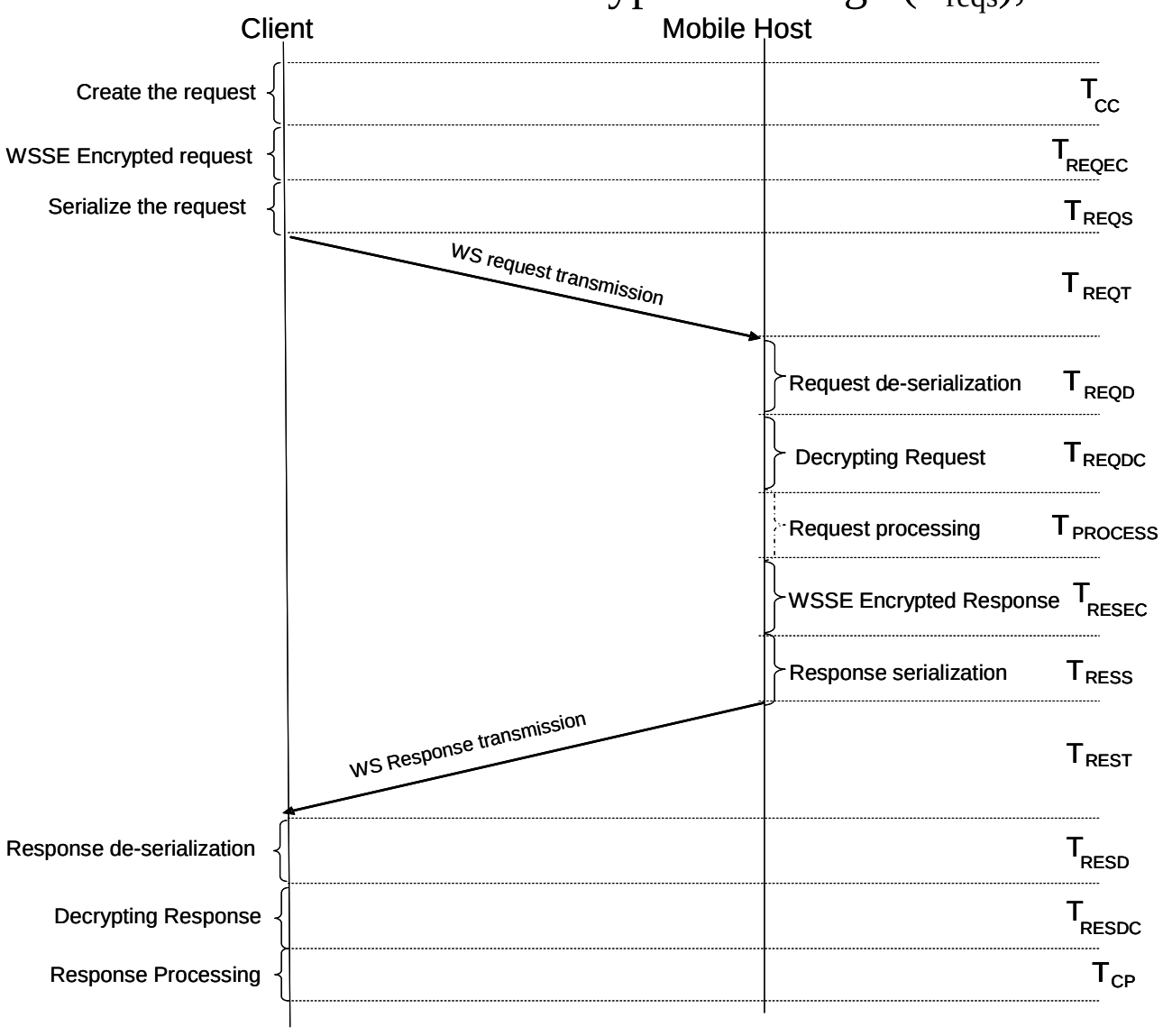

**Figure 6** *Secured mobile web service invocation: operations and time stamps*

$$T_{mwsp} = T_{cc} + T_{reqec} + T_{reqs} + T_{reqt} + T_{reqd} + T_{reqdc} + T_{process} + T_{resec} + T_{ress} + T_{rest} + T_{resd} + T_{resdc} + T_{cp} \qquad - (1)$$

The exact estimation of the $T_{reqt}$ and $T_{rest}$ time is not possible as the process needs the synchronization of time stamps of both Mobile Host and client. Moreover these transmission times were observed during our previous analysis [28]. Those results showed 90% of total invocation cycle is transmission time. So to analyze the minute extra delays due to security load, the whole invocation cycle is observed with both the invocation and processing of the WS request at the Mobile Host itself, thus eliminating the transmission aspects.

### 5.4 Evaluation Model

The main idea of our study was to realize the WS-Security standards for the mobile web service provisioning. For achieving this, different encryption algorithms, signer algorithms and authentication principles were analyzed in the Mobile Host domain. The performance of the Mobile Host was observed during the feasibility analysis, for reasonable quality of service. The parameters of interest were extra delay and variation in stability of the Mobile Host with the introduction of the security overhead. The implemented case-by-case solutions were evaluated recursively. Some of the results are discussed here:

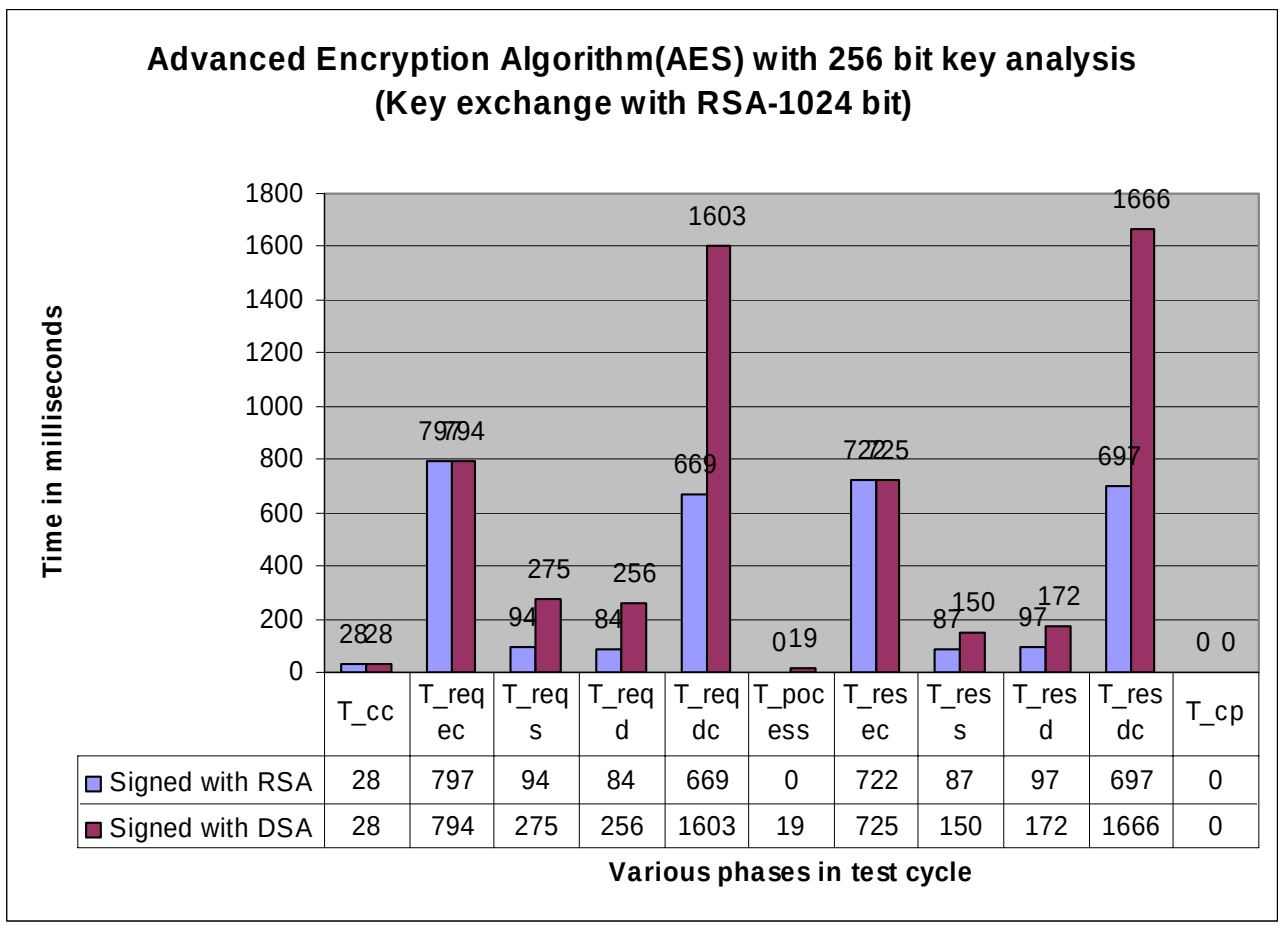

| | T_cc | T_reqec | T_reqs | T_reqd | T_reqdc | T_pocess | T_resec | T_ress | T_resd | T_resdc | T_cp |
|---|---|---|---|---|---|---|---|---|---|---|---|
| Signed with RSA | 28 | 797 | 94 | 84 | 669 | 0 | 722 | 87 | 97 | 697 | 0 |
| Signed with DSA | 28 | 794 | 275 | 256 | 1603 | 19 | 725 | 150 | 172 | 1666 | 0 |

**Figure 7** *Timestamps of various phases of a message-level secured web service cycle*

Figure 7 depicts times taken for various phases of a message level secured web service request/response cycle. The original message was ciphered with AES-256 algorithm and its key is exchanged with RSA-1024 PKI algorithm. To summarize further, the request message was 1 KB and response message was 2 KB. The total cycle for highly secured communication, AES-256 bit ciphered, cost around ~3 sec with RSAwithSHA1 signature and ~5.5 sec for DSAwithSHA1 signature.

From the performance model, we can derive the mobile WS message security effort ($T_{mwsse}$) as follows:

$$T_{mwsse} \sim= T_{reqec} + T_{reqdc} + T_{resec} + T_{resdc} \qquad - (2)$$

The extra load to the message size caused by the added security information and the extra delay thus obtained are not of the main concern as this all adds to the transmission delay. With the advent of the interim-generation technologies like GPRS [29] and EDGE [30], and third-generation technologies like UMTS [31], still higher data transmission rates are achieved in the wireless domain, in the order of few hundreds of Kbs to 2 Mbs. Most recently with the advent of 4G technologies and their deployment in south Asian countries suggests that mobile data transmissions of the rate of few GB is also possible [32]. These developments should drastically reduce the transmission delays and thus make the Mobile Host soon realizable in commercial environments also.

Based on this, we can say that the additional efforts as shown in figure 7, in achieving the highest possible secured web service communication, are reasonable. The following feasibility report further shows the message-level security analysis against various symmetric algorithms for an entire request-response web service cycle.

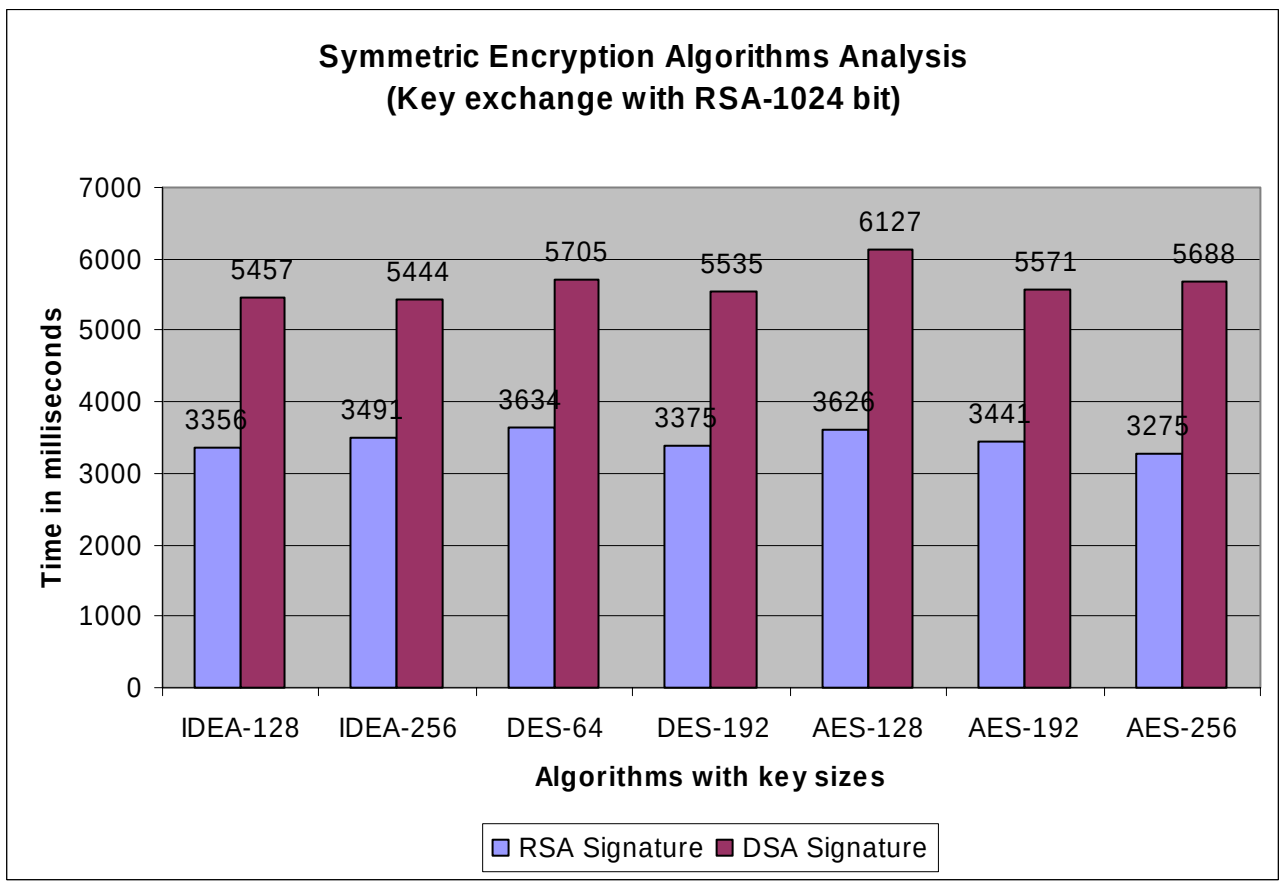

**Figure 8** *Comparison of timestamps with various symmetric key algorithms*

The analysis shown in Figure 8 emphasizes that not much effort difference exists on security front, out of mandatory standards. We can conclude that the best way of securing messages in mobile web service provisioning is to use AES symmetric encryption with 256 bit key, RSA 1024 bit key exchange mechanism and RSAwithSHA1 signature.

## 6. CONCLUSION AND FUTURE WORK

This paper mainly discussed and analyzed the security issues of mobile web services. First we have introduced mobile web service provisioning and later discussed its problems and aspects with security and tried to give the best possible message-level security scenario for mobile web services. Our future research in this domain includes providing proper end-point security for the Mobile Host using Single-sign-on with SAML and LA standards. We also want to have a detailed performance analysis of the Mobile Host with full security features through real-time applications.

Based on our till-date realization on security awareness, we conclude that secure web service provisioning in mobile networks is a great challenge. And as the mechanisms developed for traditional networks are not always appropriate for the mobile environment, this area still holds ample room for further research.

## ACKNOWLEDGEMENT

The work is supported by German Research Foundation (DFG) as part of the Graduate School ”Software for Mobile Communication Systems” at RWTH Aachen University. The authors would also like to thank M. Gerdes and R. Levenshteyn of Ericsson Research for their help and support.